\documentclass[12pt]{article}
\usepackage{amsmath,amssymb,epsfig,color}
\usepackage{feynmf}
\usepackage{psfrag}
\usepackage{graphicx}
\usepackage{bm}

\newcommand{\be}{\begin{equation}}  
\newcommand{\ee}{\end{equation}} 
\def\slash#1{#1\!\!\!/\!\,\,}  
\newcommand{\nl}{\nonumber \\ }

\newcommand{\order}{{\cal O}}

\def\Dslash{\rlap{\hspace{0.07cm}/}{D}}

\newcommand*\oline[1]{%
  \vbox{%
    \hrule height 0.5pt
    \kern0.25ex
    \hbox{%
      \kern-0.1em
      \ifmmode#1\else\ensuremath{#1}\fi
      \kern-0.1em
    }
  }
}

\allowdisplaybreaks

\begin{document}

\begin{titlepage}

\begin{flushright}
EFI Preprint 12-8\\
WSU-HEP-1204 \\
December 18, 2012
\end{flushright}

\vspace{0.7cm}
\begin{center}
\Large\bf 
The NRQED Lagrangian at order $1/M^4$
\end{center}

\vspace{0.8cm}
\begin{center}
{\sc   Richard J. Hill$^{(a)}$, Gabriel Lee$^{(a)}$, Gil Paz$^{(b)}$ and Mikhail P. Solon$^{(a)}$ } \\
\vspace{0.4cm}
{\it 
$^{(a)}$ Enrico Fermi Institute and Department of Physics \\
University of Chicago, Chicago, Illinois 60637, USA
}\\
\vspace{0.7cm}
{\it 
$^{(b)}$ 
Department of Physics and Astronomy \\
Wayne State University, Detroit, Michigan 48201, USA 
}
\end{center}
\vspace{1.0cm}

\begin{abstract}
\vspace{0.2cm}
\noindent  The parity and time-reversal invariant effective Lagrangian
for a heavy fermion interacting with an Abelian gauge field, i.e.,
NRQED, is constructed through order $1/M^4$. The implementation of
Lorentz invariance in the effective theory becomes nontrivial at this order,  
and a complete solution for Wilson coefficient constraints is obtained.   Matching
conditions in the one-fermion sector are presented in terms of form
factors and two-photon matrix elements of the nucleon.  
The extension of NRQED to describe interactions of the heavy fermion with a
light fermion is introduced. 
Sample applications are discussed; these include
the computation of nuclear structure effects in atomic bound
states,  the model-independent analysis of radiative corrections to
low-energy  lepton-nucleon scattering, and  the study of static
electromagnetic properties of nucleons.  

\end{abstract}
\vfil

\end{titlepage}

\section{Introduction}

Nonrelativistic QED (NRQED) is an effective field theory~\cite{Caswell:1985ui} 
describing the interactions of nonrelativistic fermions with electromagnetic fields.   
NRQED interactions at order $1/M^4$ have become relevant for  
describing radiative corrections to proton structure contributions in 
hydrogenic bound state spectroscopy~\cite{Pohl:2010zza,Hill:2011wy}.  
The NRQED Lagrangian, properly constrained by Lorentz invariance,
trivializes the derivation of low-energy theorems of Compton scattering~\cite{Ragusa:1993rm}
and automatically incorporates the intricate
singularity structure of scattering amplitudes~\cite{Bardeen:1969aw,Tarrach:1975tu}.
It can be used to rigorously compute radiative corrections to low-energy lepton-nucleon 
scattering, and it also provides a model-independent framework within which to analyze static 
properties of nucleons, such as polarizabilities and generalized electromagnetic moments~\cite{Lvov:1993fp}. 
In this paper we derive a complete basis of operators and coefficient constraints 
through order $1/M^4$ for the effective theory of nonrelativistic nucleons and 
leptons interacting with photons.%
\footnote{
For definiteness we will often refer to the heavy fermion $\psi$ as the ``nucleon,'' 
and to a second fermion $\chi$ in Section \ref{sec:fourfermion} as the ``lepton.''    
} 

An important formal issue first arises at order $1/M^4$.  
As recently discussed in \cite{Heinonen:2012km}, a ``reparametrization invariance'' 
ansatz for enforcing relativistic invariance breaks down at this order. 
We derive the correct implementation of Lorentz invariance constraints and the
resulting Wilson coefficient relations through order $1/M^4$. 
For applications involving NRQED of the proton or neutron
we perform leading order (in $\alpha$) 
matching computations to relate the remaining undetermined coefficients 
to observables of lepton-nucleon scattering.  

The remainder of the paper is structured as follows.  
In Sec.~\ref{sec:abelian} we construct the 
NRQED Lagrangian in the one-fermion sector through order $1/M^4$.   This Lagrangian 
describes the interaction of a nucleon with arbitrary background electromagnetic fields.   
Sec.~\ref{sec:reparam} enforces constraints on the Wilson coefficients
deriving from relativistic invariance, first  by a variational calculation and 
then by an equivalent  invariant operator construction. 
In Sec.~\ref{sec:matching} we perform 
matching calculations in the one-fermion sector, 
relating the undetermined Wilson coefficients to form factors and observables of 
(virtual) Compton scattering. 
In Sec.~\ref{sec:fourfermion} we complete the basis of operators in the zero-fermion 
and two-fermion sectors, required to describe a proton 
interacting with a nonrelativistic lepton and dynamical photon.  
For applications to electron-proton scattering ($m_e, E \ll M$) we 
also consider the extension to the case of a relativistic 
lepton.  
Sec.~\ref{sec:apps} provides a brief discussion of applications to 
the computation of nuclear structure effects in atomic bound states;  
to the model-independent analysis of radiative corrections
to low-energy  lepton-nucleon scattering; and to 
the study of static electromagnetic properties of nucleons. 
Sec.~\ref{sec:summary} provides a concluding discussion. 

\section{Lagrangian \label{sec:abelian}}

Consider the Lagrangian for a heavy fermion coupled to an Abelian gauge field. 
We enforce Hermiticity and invariance under  
parity, time-reversal and rotational symmetries. 
We also perform field redefinitions
to eliminate time derivatives acting on the fermion field (apart from the leading term);
we refer to this choice as the ``canonical form'' of the heavy particle Lagrangian. 
We thus find in the one-fermion sector through $1/M^4$, 
\begin{align}\label{eq:abelian}
  {\cal L} &= \psi^\dagger
  \bigg\{  i D_t  + c_2 {\bm{D}^2 \over 2M}  + c_4 {\bm{D}^4 \over 8 M^3} +
  c_F g{ \bm{\sigma}\cdot \bm{B} \over 2M}   
+ c_D g{ [\bm{\partial}\cdot \bm{E}] \over 8 M^2}  
+ i c_S g{ \bm{\sigma}
    \cdot ( \bm{D} \times \bm{E} - \bm{E}\times \bm{D} ) \over 8 M^2} 
\nl
&\quad
+ c_{W1}g {  \{ \bm{D}^2 ,  \bm{\sigma}\cdot \bm{B} \}  \over 8 M^3}  
- c_{W2}g {  D^i \bm{\sigma}\cdot
    \bm{B} D^i \over 4 M^3 }  + c_{p^\prime p} g { \bm{\sigma} \cdot
    \bm{D} \bm{B}\cdot \bm{D} + \bm{D}\cdot\bm{B} \bm{\sigma}\cdot \bm{D}
    \over  8 M^3} 
\nl
&\quad 
+ i c_M g { \{ \bm{D}^i ,  [\bm{\partial} \times \bm{B}]^i \} \over 8 M^3}
+ c_{A1} g^2{ \bm{B}^2 - \bm{E}^2 \over 8 M^3}  - c_{A2} g^2{ \bm{E}^2 \over 16 M^3 } 
\nl
&\quad 
+ c_{X1}g { [ \bm{D}^2 , \bm{D}\cdot \bm{E} + \bm{E}\cdot\bm{D} ] \over M^4 }
+ c_{X2}g { \{ \bm{D}^2 , [\bm{\partial}\cdot\bm{E}] \} \over M^4 }
+ c_{X3}g { [\bm{\partial}^2 \bm{\partial}\cdot\bm{E}] \over M^4 } 
\nl
&\quad 
+ i c_{X4}g^2 { \{ \bm{D}^i , [\bm{E}\times\bm{B}]^i \} \over M^4 } 
+ ic_{X5} g { D^i \bm{\sigma}\cdot ( \bm{D}\times\bm{E} - \bm{E}\times\bm{D} )D^i   \over M^4} 
+ ic_{X6} g { \epsilon^{ijk} \sigma^i D^j [\bm{\partial}\cdot\bm{E}] D^k \over M^4} 
\nl
&\quad
+ c_{X7} g^2 { \bm{\sigma}\cdot\bm{B} [\bm{\partial}\cdot\bm{E}] \over M^4} 
+ c_{X8} g^2 { [\bm{E}\cdot\bm{\partial} \bm{\sigma}\cdot\bm{B} ] \over M^4}
+ c_{X9} g^2 { [\bm{B}\cdot\bm{\partial} \bm{\sigma}\cdot\bm{E} ] \over M^4} 
+ c_{X10} g^2 { [E^i \bm{\sigma}\cdot\bm{\partial} B^i] \over M^4}
\nl
&\quad
+ c_{X11} g^2 { [B^i \bm{\sigma}\cdot\bm{\partial} E^i] \over M^4} 
+ c_{X12} g^2 { \bm{\sigma}\cdot \bm{E}\times [{\partial_t}\bm{E}-\bm{\partial}\times\bm{B} ] \over M^4} 
+ \mathcal{O}(1/M^5)
 \bigg\} \psi  \,.
\end{align}
We have defined $D_t = {\partial/\partial t} + i g Z A^0$, $D^i = {\partial/\partial x^i}-ig Z A^i$, 
where $-gZ = -e$, $+e$ or $0$ for an electron, proton or neutron, respectively. 
We use the summation convention $X^i Y^i \equiv \sum_{i=1}^3 X^i Y^i$, 
and define $[X,Y]\equiv XY-YX$, $\{X,Y\} \equiv XY+YX$ to denote commutators and 
anticommutators as usual.  Square brackets around quantities imply that derivatives
act only within the bracket.  
Electric and magnetic fields are defined as usual by $\bm{E}=-[\partial_t \bm{A}] - [\bm{\partial}A^0]$
and $\bm{B}=[\bm{\partial}\times\bm{A}]$. 
By the definition of $\bm{E}$ and $\bm{B}$, 
$[\bm{\partial}\cdot\bm{B}]=0$ and $[\partial_t \bm{B}+\bm{\partial}\times\bm{E}]=0$. 

The most general term in (\ref{eq:abelian}) is obtained
by constructing all possible rotationally invariant, Hermitian combinations 
of $iD_t$, $D^i$, $E^i$, $iB^i$, $i\sigma^i$, with parity requiring an even number
of factors of $D^i$ and $E^i$. The operators through 
$1/M^3$ were previously introduced in~\cite{Caswell:1985ui,Kinoshita:1995mt,Manohar:1997qy}.
Terms at $1/M^4$ with two field strength factors $E^i$ or $B^i$ 
are straightforward to tabulate; 
note that we have used $[\partial_t \bm{B}] = -[\bm{\partial}\times\bm{E}]$ 
and the assumption of canonical form to eliminate time derivatives of the magnetic field.  
Remaining terms at $1/M^4$ involve one factor of 
electric field $E^i$ and three spatial derivatives $D^i$.   
Spin-independent terms are straightforward to tabulate; the basis of operators parametrized 
by $c_{X1}$, $c_{X2}$, $c_{X3}$ differs from other possible choices by terms involving 
commutators $[D^i,D^j]$, i.e., terms with two field strengths. 
For spin-dependent terms we use $[\bm{\partial}\times\bm{E}]=-[\partial_t \bm{B}]$
and the assumption of canonical form to eliminate occurrences of $[\bm{\partial}\times\bm{E}]$. 
The three-vector identity, 
\be
D^i (\bm{E}\times \bm{\sigma} )^j + (\bm{\sigma}\times\bm{D})^j E^i + \sigma^i (\bm{D}\times\bm{E})^j 
= \bm{D}\cdot\bm{E}\times\bm{\sigma} \delta^{ij} \,,
\ee
applied to remaining terms of the form $\psi^\dagger D^i (\dots )D^j \psi$, leaves 
the basis of operators parametrized by $c_{X5}$, $c_{X6}$.

\section{Relativistic invariance\label{sec:reparam}}

The Lagrangian (\ref{eq:abelian}) is invariant, by construction, under rotations and
spacetime translations. The remaining 
constraints of relativity are enforced by
demanding invariance under boosts.  Here we derive 
these additional constraints,  first by a variational calculation in 
Sec.~\ref{sec:vari}, and then by an equivalent invariant operator construction 
in Sec.~\ref{sec:invariant}. 

\subsection{Variational method \label{sec:vari}}

As detailed in \cite{Heinonen:2012km}, under infinitesimal boosts, 
with infinitesimal boost parameter $\bm{\eta} = - \bm{q}/M$, 
we may choose the heavy fermion to transform as
\begin{align}\label{eq:lorentz}
\psi \to & e^{-i\bm{q}\cdot\bm{x}}\left\{  
1 + {i\bm{q}\cdot\bm{D}\over 2 M^2} + {i \bm{q}\cdot\bm{D} \bm{D}^2 \over 4 M^4} 
- 
{ \bm{\sigma}\times \bm{q} \cdot \bm{D} \over 4 M^2} 
\left[ 1 + {\bm{D}^2\over 4 M^2} \right]  \right. \nl
& \quad \quad \quad \quad \quad \quad \quad \quad \left. + {i c_Dg \over 8 M^3} \bm{q}\cdot \bm{E} 
+ {c_Sg \over 8 M^3} \bm{q}\cdot\bm{\sigma}\times\bm{E} + {\cal O}(g/M^4, 1/M^6) + \dots
\right\}
\psi \,,
\end{align}
while derivatives and gauge fields transform as Lorentz vectors, so that
\begin{align} \label{eq:lorentz2}
D_t \to D_t + {1\over M} \bm{q}\cdot\bm{D} \,, \quad
\bm{D} \to \bm{D} + {1\over M} \bm{q} D_t \,, \quad
\bm{E} \to \bm{E} + {1\over M} \bm{q}\times \bm{B} \,, \quad
\bm{B} &\to \bm{B} - {1\over M} \bm{q}\times \bm{E} \,.
\end{align} 
Field strength-dependent terms in (\ref{eq:lorentz}) have been chosen to 
maintain canonical form.  Since we are interested in the canonical Lagrangian
through order $1/M^4$, we need not specify the explicit form of the 
order $1/M^4$ field strength-dependent terms, 
denoted by ${\cal O}(g/M^4)$.  A straightforward computation yields 
\be
\delta {\cal L} = {1\over M} \delta {\cal L}_1 + {1\over M^2} \delta {\cal L}_2 
+  {1\over M^3} \delta {\cal L}_3
+ {1\over M^4} \delta {\cal L}_4 
+ \dots \,,
\ee
where
\begin{align} 
\delta {\cal L}_1 &= \psi^\dagger \left[  (1- c_2) i\bm{q}\cdot \bm{D}  \right] \psi \,, 
\nl
\delta {\cal L}_2 &= \psi^\dagger \left[ 
-\frac12(1-c_2) \{ \bm{q}\cdot\bm{D} , D_t \} 
+ \frac{g}{4} ( Z - 2 c_F + c_S ) \bm{\sigma}\times \bm{q} \cdot \bm{E} 
   \right] \psi \,, \nl
\delta {\cal L}_3 &= \psi^\dagger \bigg[ 
\frac{g}{8} \bm{q}\cdot [\bm{\partial}\times \bm{B}] \left( c_F - c_D + 2 c_M \right) 
+ {i\over 4} \{ \bm{q}\cdot\bm{D}, \bm{D}^2 \} \left( c_2 - c_4 \right) 
\nl
&\quad
+ {ig\over 8}\{\bm{q}\cdot \bm{D} , \bm{\sigma}\cdot \bm{B} \} 
\left( c_2 Z  +2c_F - c_S -2c_{W1} + 2c_{W2}  \right)
\nl
&\quad
+ {ig\over 8}\{\bm{\sigma}\cdot \bm{D} , \bm{q}\cdot \bm{B} \} 
\left( -c_2Z + c_F - c_{p^\prime p} \right)
+ {ig\over 8} \bm{q}\cdot\bm{\sigma}(\bm{D}\cdot\bm{B}+\bm{B}\cdot\bm{D}) 
\left(-c_F + c_S - c_{p^\prime p} \right)
\bigg] \psi .
\end{align} 
From $\delta {\cal L}_1$, $\delta {\cal L}_2$ and $\delta {\cal L}_3$, we 
find~\cite{Manohar:1997qy,Hill:2011wy}%
\footnote{
The relations in \cite{Manohar:1997qy} assume $Z=1$. As noted in \cite{Hill:2011wy}, 
we find the opposite sign in the relation for $c_M$ in (\ref{eq:constrain12}) compared 
to \cite{Manohar:1997qy}. 
}
\begin{align}\label{eq:constrain12} 
c_2 &= 1 \,, \quad
 c_S = 2 c_F - Z \,,\quad
  c_4 = 1 \,, \quad
2c_M = c_D - c_F \,, \quad
c_{W2} = c_{W1} - Z \,, \quad 
c_{p^\prime p} = c_F - Z \,.  
\end{align}
Employing the above relations, 
the variation $\delta {\cal L}_4$ takes the form 
\begin{align}
\delta {\cal L}_4 &= \psi^{\dagger} \bigg[ 
 {ig\over 8} [\bm{D}^2 , \bm{q}\cdot\bm{E} ] \left( \frac{5Z}{4} - c_F + c_D - 32 c_{X1} \right)
+ {ig\over 8}  \{ \bm{q}\cdot\bm{D}, [\bm{\partial}\cdot\bm{E}] \} 
\left( 
-\frac{Z}{4} + c_F - 16 c_{X2} 
\right)
\nl
&\quad 
+ {g^2\over 8}  \bm{q}\cdot\bm{E}\times\bm{B} 
\left( 
\frac{Z^2}{2} + 2c_F(Z-c_F) - 2 Z c_D + c_{A2} + 16 c_{X4} 
\right)
\nl
&\quad
+ \frac{g}{8} [ \bm{q}\cdot \bm{\sigma}\times \bm{\partial} \bm{\partial}\cdot \bm{E} ]
\left( -Z + c_F - \frac14 c_D + c_{W1} + 8 c_{X6} \right)
\nl
&\quad
+ \frac{g}{8} D^i \left( q^i (\bm{E}\times \bm{\sigma})^j +  
(\bm{E}\times \bm{\sigma})^i q^j 
+ \bm{\sigma}\times\bm{q}\cdot\bm{E} \delta^{ij} \right) D^j \left( \frac{Z}{2} - 2 c_F + 16 c_{X5} \right) 
\bigg] \psi \,,
\end{align}
where we have suppressed terms that are removed by field strength-dependent 
modifications of the boost generator, denoted by ${\cal O} (g/M^4)$ in  (\ref{eq:lorentz}).
We readily find
\begin{align}\label{eq:constrain4} 
32 c_{X1} &= \frac{5Z}{4} - c_F + c_D \,, \nl
32 c_{X2} &= -\frac{Z}{2} + 2 c_F \,, \nl
32 c_{X4} &= -Z^2 - 4 c_F(Z-c_F) + 4 Z c_D - 2 c_{A2} \,, \nl
32 c_{X5} &= -Z + 4 c_F \,, \nl
32 c_{X6} &= 4(Z - c_F) + c_D -4 c_{W1} \,, 
\end{align}
while coefficients $c_{X3}$ and $c_{X7\dots X12}$ are not constrained by Lorentz invariance. 
We thus find that seven new quantities are required at order $1/M^4$ to 
describe the proton's response to arbitrary background electromagnetic fields.   

\subsection{Invariant operators \label{sec:invariant}} 

An alternate method for enforcing Lorentz invariance is to construct the 
Lagrangian from explicitly invariant operators. We summarize here the main points; 
for details see \cite{Heinonen:2012km}. 

The basic building block in the construction is the field $\Psi_v = \Gamma(v,iD) \psi_v$,
where $\psi_v$ is a Dirac spinor field with $\slash{v} \psi_v = \psi_v$. 
The matrix-valued operator $\Gamma(v,iD)$ is determined such that 
under an infinitesimal boost $\Lambda$, where 
$\Lambda^\mu_{\,\,\nu}v^\nu = v^\mu + q^\mu/M$,  
the field $\Psi_v$ has a simple transformation law: $\Psi_v \to
e^{iq\cdot x} \Psi_v$. 
Noting that $e^{-iq\cdot x} (iD^\mu + M v^\mu + q^\mu )
e^{iq\cdot x} = iD^\mu + M v^\mu$, we may thus build invariant bilinears from
contractions of polynomials of $\gamma^\mu$ and ${\cal V}^\mu \equiv v^\mu
+ iD^\mu/M $, between $\oline{\Psi}_v$ and $\Psi_v$. 

The function $\Gamma(v,iD)$ is a solution to the invariance equation, 
\be
\Gamma(v+q/M,iD-q) \Lambda^{-1} W(\Lambda, iD+Mv) = \Gamma(v,iD),
\ee
where $W(\Lambda,p)$ is 
an element of the little group for timelike invariant vector $v^\mu$, following from the 
theory of induced representations of the Lorentz group.   
Up to the relevant order for determining the $1/M^4$ Lagrangian we have~\cite{Heinonen:2012km}
\begin{align} \label{eq:Gamma}
\Gamma = 1 +& 
{i \Dslash_\perp \over 2M}
+ \frac1{M^2} \left\{ -\frac18 (iD_\perp)^2
-\frac12 i \Dslash_\perp i v \cdot D\right\}  
+ \frac1{M^3}
\left\{  \frac14 (iD_\perp)^2 i v \cdot D 
\right.
\nl & \left.
+ \frac{i
\Dslash_\perp}{2} \left[ -\frac38 (iD_\perp)^2 + (iv \cdot D)^2
\right] 
+\frac{gZ}{8} F_{\mu \nu} v^{\mu} D_\perp^\nu 
+ \frac{gZ}{16} \sigma_\perp^{\mu \nu} F_{\mu \nu} i \Dslash_\perp \right\}  +
\dots,
\end{align} 
where we have defined $D_\perp^\mu \equiv D^\mu - v^\mu v \cdot D$, and for Abelian gauge fields 
$F_{\mu\nu}\equiv \partial_\mu A_\nu - \partial_\nu A_\mu$.  
Note that the last two terms of (\ref{eq:Gamma}) 
are absent in the ansatz for reparametrization
invariance given in \cite{Luke:1992cs}, leading to incorrect Lorentz
invariance constraints at $1/M^4$ and beyond. This subtlety is
explained in \cite{Heinonen:2012km}.

A complete basis of invariant bilinears required through order $1/M^4$
is
\begin{align}
{\cal L} &= \oline{\Psi}_v \bigg\{ 
M(\slash{{\cal V} }-1) 
- a_F g {\sigma^{\mu\nu} F_{\mu\nu} \over 4 M}  
+ i a_D g { \{ {\cal V}_\mu , [M {\cal V}_\nu, F^{\mu\nu} ] \} \over 16 M^2} 
- a_{W1} g { [M{\cal V}^\alpha, [M{\cal V}_\alpha, \sigma^{\mu\nu} F_{\mu\nu} ] ] \over 16 M^3} 
\nl
&\quad
+ a_{A1} g^2 { F_{\mu\nu}F^{\mu\nu} \over 16 M^3} 
+ a_{A2} g^2 { {\cal V}_\alpha F^{\mu\alpha} F_{\mu\beta} {\cal V}^\beta \over 16 M^3}  
\bigg\}\Psi_v 
+ a_{X3} {\cal B}_{X3} + \sum_{i=7}^{12} a_{Xi} {\cal B}_{Xi} .
\end{align}
The bilinears ${\cal B}_{Xi}$ for $i=3, 7\dots 12$ 
are chosen to reduce to the respective operators multiplying $c_{Xi}$ in (\ref{eq:abelian}) 
upon setting $v^\mu=(1,0,0,0)$ and neglecting $1/M$ suppressed corrections. 
Since we are concerned only with the Lagrangian through order $1/M^4$ 
we do not specify an explicit choice for these ${\cal B}_{Xi}$. 
A computation shows that the field redefinition to recover 
canonical form is
\begin{align}\label{eq:psi}
\psi_v &=\bigg\{ 1 + {1 \over 4 M^2}(iD_\perp)^2 \left(1 - {i v\cdot D\over M} \right) 
- \frac{gZ}{16M^2}\sigma_\perp^{\mu \nu} F_{\mu \nu} 
- \frac{gZ}{4M^3} D_\perp^\mu v^\alpha F_{\alpha \mu} 
+ \frac{igZ}{4M^3}\sigma_{\mu \nu} D_\perp^\mu v_\alpha F^{\alpha \nu}  
\nl
&\quad 
-\frac{gZ}{8M^3} v^\alpha F_{\alpha \mu} D_\perp^\mu 
+ {g a_F\over 4 M^3} \left[ -D_\perp^\mu v^\alpha F_{\alpha \mu} + i \sigma_{\mu \nu} D_\perp^\mu v_\alpha F^{\alpha \nu} \right] 
- {g a_D\over 8 M^3} v^\alpha F_{\alpha \mu} D_\perp^\mu  
\nl
 &
\quad
+\frac{ig a_{W1}}{8M^3} \sigma_{\mu \nu}[ D_\perp^\mu, v_\alpha F^{\alpha \nu}]  \bigg\} \psi_v^{\prime} \, .
\end{align}
Upon setting $v^\mu=(1,0,0,0)$, 
the resulting Lagrangian, expressed in terms of $\psi_v^{\prime}$, 
is identical to the previous result (\ref{eq:abelian}) with
constraints (\ref{eq:constrain12}) and (\ref{eq:constrain4}).

\section{Matching: one-fermion sector \label{sec:matching}}

In contrast to NRQED for the electron or other fundamental fermions, 
the matching for a composite particle such as the proton cannot be performed 
perturbatively.   We instead must appeal to nonperturbative (e.g. lattice) 
methods, or to experimental measurements.  This section relates the matching
conditions in the one-fermion sector 
to standard form factors and two-photon matrix elements of the nucleon. 

\subsection{One-photon matching}

\begin{figure}[htb]
\begin{center}
\includegraphics[scale=0.9]{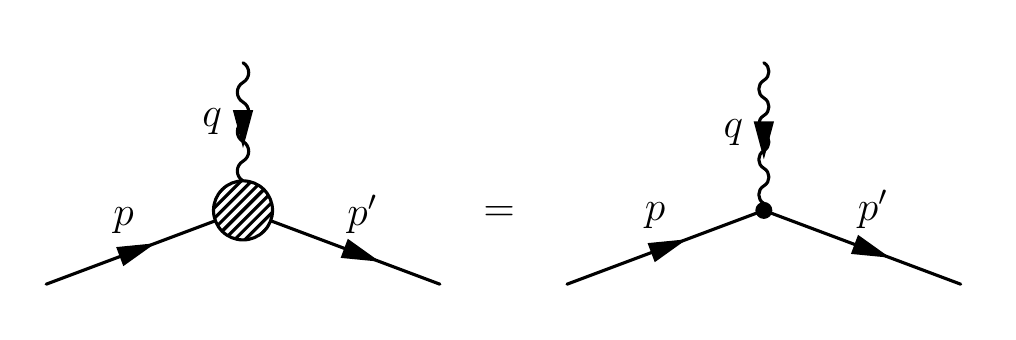}
\end{center}
\vspace{-5mm}
\caption{\label{fig:one_photon}
Tree level matching of the one-photon 
amplitude in the full theory and NRQED. 
The black dot in the diagram on the right-hand side 
represents single-photon NRQED vertices.
}
\end{figure}

Consider first the operators contributing to the one-photon
matrix element of the nucleon.   The matching is performed in terms of 
standard invariant form factors,
\be\label{eq:FF}
\langle N(p')|J_\mu^{\rm em}|N(p)\rangle=\oline u(p')
\Gamma_\mu(q) 
u(p)\,,
\quad
\Gamma_\mu(q) \equiv \gamma_\mu F^N_1(q^2)+\frac{i\sigma_{\mu\nu}}{2 M_N}F^N_2(q^2)q^\nu \,,
\ee
where $q=p'-p$ and $N$ denotes a proton or neutron; we suppress the superscript 
$N$ in the following.   Equating the effective theory with the full theory,\footnote{
The nonrelativistic normalization of states in NRQED is obtained 
using $\bar{u}(p) u(p) = M/E_{\bm{p}}$ in (\ref{eq:FF}).  
}
we find (cf. Fig.~\ref{fig:one_photon})
\begin{align}\label{eq:c1photon}
c_F &= \bar{F}_1 +\bar{F}_2  
\equiv Z + a_N + \order(\alpha) 
\,, \nl  
c_D &= \bar{F}_1 + 2\bar{F}_2 + 8 \bar{F}_1^\prime 
\equiv Z + \frac43 M^2 (r_E^N)^2 + \order(\alpha) 
\,, \nl 
c_{W1} &= \bar{F}_1 + \frac12 \bar{F}_2 + 4 \bar{F}_1^\prime + 4 \bar{F}_2^\prime \,, \nl 
c_{X3} &= \frac18 \bar{F}_1^\prime + \frac14 \bar{F}_2^\prime + \frac12 \bar{F}_1^{\prime\prime} \,,
\end{align}
where $Z$ denotes the electric charge ($Z=1$ for the proton and $Z=0$ for the neutron), $a_N$ 
is the anomalous magnetic moment of the nucleon, and $r_E^N$ is the nucleon charge radius.
We have introduced dimensionless barred quantities to denote derivatives 
with respect to $q^2/M^2$ at $q^2=0$: $\bar{F}_1 \equiv F_1(0)= Z$, $\bar{F}_2\equiv F_2(0)= a_N$, 
$\bar{F}_i^\prime \equiv M^2 F_i^\prime(0)$, etc. The new quantity $\bar{F}_1^{\prime\prime}$ appears at $1/M^4$.
Expressions for other Wilson coefficients through $1/M^3$ in terms of form factors can be found using 
(\ref{eq:constrain12}). At $1/M^4$, we also find
\begin{align} \label{eq:c1photonXi}
c_{X1} &= \frac{5}{128} \bar{F}_1 +  \frac{1}{32}\bar{F}_2 + \frac14 \bar{F}_1^\prime \,, \nl 
c_{X2} &= \frac{3}{64} \bar{F}_1 +  \frac{1}{16} \bar{F}_2 \,, \nl
c_{X5} &= \frac{3}{32} \bar{F}_1 + \frac18 \bar{F}_2 \,, \nl
c_{X6} &= - \frac{3}{32} \bar{F}_1 - \frac18 \bar{F}_2 - \frac14 \bar{F}_1^\prime - \frac12 \bar{F}_2^\prime \,,
\end{align}
and it is readily verified that these expressions satisfy the constraints (\ref{eq:constrain4}). In the presence of radiative corrections, the form factors on the right-hand sides
of (\ref{eq:c1photon}) and (\ref{eq:c1photonXi}) should be interpreted in an appropriate infrared regularization scheme; 
alternatively, the matching may be performed with infrared finite observables. 
The corresponding infrared subtractions and ultraviolet renormalizations must be 
performed to obtain the Wilson coefficients including radiative corrections.%
\footnote{The expressions on the right-hand sides of (\ref{eq:c1photon}) and (\ref{eq:c1photonXi}) 
correspond to those referred to as $c_i^{\rm QED}$ 
in \cite{Kinoshita:1995mt}.  The renormalization procedure in dimensional regularization 
is described in \cite{Manohar:1997qy}. 
}

\subsection{Two-photon matching} 

\begin{figure}[htb]
\begin{center}
\includegraphics[scale=0.9]{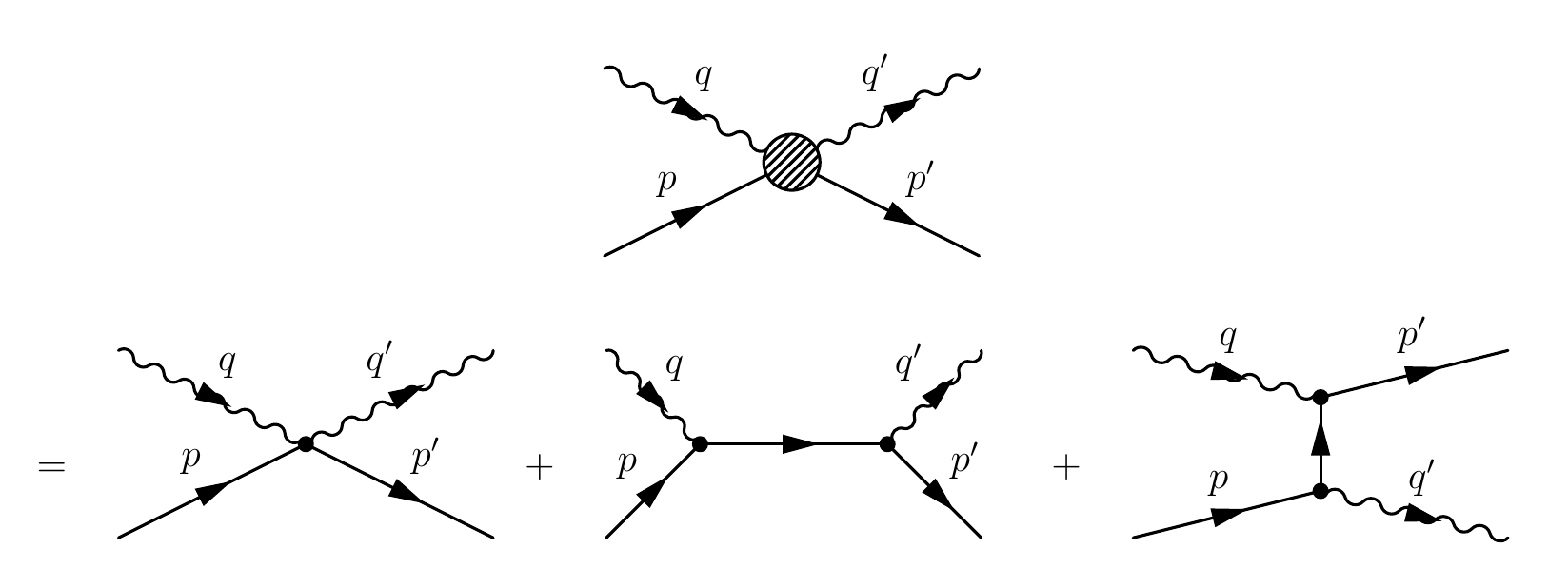}
\end{center}
\vspace{-5mm}
\caption{\label{fig:two_photon}
Tree-level matching of the Compton scattering amplitude
in the full theory and NRQED. 
The black vertices in the diagrams on the right-hand side 
represent NRQED vertices. 
}
\end{figure}

The Compton scattering process, 
$\gamma^*(q) N(p) \to \gamma(q^\prime) N(p^\prime)$, 
with one virtual and one real photon is sufficient 
to determine the remaining coefficients in the $1/M^4$ NRQED Lagrangian. 
Consider the low-energy expansion of the virtual Compton scattering amplitude
as depicted in Fig.~\ref{fig:two_photon}. 
Let us define a conventional separation
\be
{\cal M}^{\mu\nu} \equiv {\cal M}_{\rm Born}^{\mu\nu} + {\cal M}_{\text{non-Born}}^{\mu\nu} \,,
\ee
by declaring that ``Born'' terms are 
defined by the ``sticking in form factors'' prescription 
using the form factors of (\ref{eq:FF}).  Explicitly, 
\be\label{eq:siff}
{\cal M}^{\mu\nu}_{\rm Born} 
\equiv - g^2 \bar{u}(p^\prime) \bigg\{ 
\Gamma^\nu(-q^\prime) 
{1\over \slash{p}+\slash{q} - M} 
\Gamma^\mu(q) 
+ \Gamma^\mu(q) 
{1\over \slash{p}-\slash{q}^\prime - M} 
\Gamma^\nu(-q^\prime) 
\bigg\} u(p) \,. 
\ee
This convention ensures 
that $q_\mu {\cal M}_{\rm Born}^{\mu\nu} = q^\prime_\nu {\cal M}_{\rm Born}^{\mu\nu}=0$, 
so that the same Ward identities, 
$q_\mu {\cal M}_{\text{non-Born}}^{\mu\nu} = q^\prime_\nu {\cal M}_{\text{non-Born}}^{\mu\nu}=0$, 
may be applied to constrain ${\cal M}_{\text{non-Born}}^{\mu\nu}$. 
We adopt a convenient basis for ${\cal M}_{\text{non-Born}}^{\mu\nu}$ as given by Drechsel et al 
[Eq.~(A10) of~\cite{Drechsel:1997xv}], 
\be
{\cal M}_{\text{non-Born}}^{\mu\nu} = \sum_{i=1}^{12} f_i(q^2, q\cdot q^\prime, q\cdot p) \rho_i^{\mu\nu} \,. 
\ee
Subtraction of the Born terms ensures that the 
residual contributions $f_i(q^2,q\cdot q^\prime, q\cdot p)$ are 
free of kinematic singularities, and may thus be Taylor 
expanded at small photon energy. 
Employing $q\cdot p \sim \omega$, $q^2 \sim \omega^2$,  $q\cdot q^\prime \sim \omega^2$, 
to the relevant order, we require 
\begin{align}\label{eq:fexpand}
f_{i}(q^2,q\cdot q^\prime, q\cdot p) &\equiv f_{i,0} + \order(\omega^2) \,, \quad i=1,2,5,6,11,12 \,,  
\nl
f_{i}(q^2,q\cdot q^\prime, q\cdot p) &\equiv f_{i,1} \, q\cdot p + \order (\omega^3) \,, \quad i=4,10 \,,
\end{align}
where we adopt the notation of \cite{Drechsel:1998zm}. 

The matching can be performed in arbitrary reference frame, 
with the results,%
\footnote{We have performed the computation both in the laboratory frame, 
$\bm{p}=0$, and in the center of mass frame, $\bm{q}+\bm{p}=\bm{0}$, by 
matching both the full theory and the effective theory to the twelve independent 
spin structures for the virtual Compton scattering process. 
}  
\begin{align} \label{eq:virtual}
c_{A1} &= \bar{F}_1^2 + 4 \bar{f}_{1,0} 
\equiv Z^2 + {4M^3\over \alpha }\beta_M + \order(\alpha) 
\,, \nl
c_{A2} &= 4 \bar{F}_1 \bar{F}_2 + 2 \bar{F}_2^2 + 16 \bar{F}_1 \bar{F}_1^\prime + 32 \bar{f}_{2,0} 
\equiv 2 a_N^2 + {8Z\over 3} M^2 (r_E^N)^2  - {8 M^3\over \alpha} ( \alpha_E + \beta_M ) + \order(\alpha) 
\,, \nl
c_{X7} &= {1\over 4} \bar{F}_1 \bar{F}_1^\prime + {1\over 2} \bar{F}_1 \bar{F}_2^\prime 
-\frac12\bar{f}_{5,0} - 2\bar{f}_{11,0} - 2\bar{f}_{12,0} \,, \nl
c_{X8} &= -{9\over 32}\bar{F}_1^2 -{3\over 8}\bar{F}_1\bar{F}_2 
-{1\over 16} \bar{F}_2^2 - \left(\bar{F}_1 +\frac12 \bar{F}_2 \right)  \bar{F}_1^\prime 
-{1\over 2} \bar{F}_1 \bar{F}_2^\prime - 4\bar{f}_{4,1} + 4 \bar{f}_{6,0} - 2\bar{f}_{10,1} \,, \nl
c_{X9} &= {5\over 64}\bar{F}_1^2 + {3\over 16} \bar{F}_1\bar{F}_2 
+{1\over 8}\bar{F}_2^2 + {1\over 2}( \bar{F}_1+\bar{F}_2) ( \bar{F}_1^\prime 
+ \bar{F}_2^\prime) - 2 \bar{f}_{10,1} \,, \nl
c_{X10} &= {7\over 32}\bar{F}_1^2 + {3\over 8} \bar{F}_1\bar{F}_2 
+ {1\over 16} \bar{F}_2^2 + 4 \bar{f}_{4,1} + 2\bar{f}_{10,1} + 2 \bar{f}_{11,0} \,, \nl
c_{X11} &= {7\over 64}\bar{F}_1^2 + {3\over 16} \bar{F}_1\bar{F}_2   
- {1\over 2}( \bar{F}_1+\bar{F}_2 ) ( \bar{F}_1^\prime + \bar{F}_2^\prime) + 2\bar{f}_{10,1} + 2\bar{f}_{11,0} \,, \nl
c_{X12} &= -{1\over 16}\bar{F}_1^2 -{1\over 8} \bar{F}_1 \bar{F}_2 -{1\over 16} \bar{F}_2^2  
- {1\over 2}( \bar{F}_1+\bar{F}_2)  \bar{F}_1^\prime
- {1\over 2}\bar{F}_1 \bar{F}_2^\prime + 4\bar{f}_{4,1} + \frac12\bar{f}_{5,0} + 2\bar{f}_{10,1} + 2\bar{f}_{11,0} \,,
\end{align}
where the dimensionless form factor derivatives, $\bar{F}^{(n)}_i$, have been introduced after (\ref{eq:c1photon}) 
and we have similarly defined dimensionless quantities 
$\bar{f}_1 = M^3 f_1$, $\bar{f}_2 = M^5 f_2$, $\bar{f}_4 = M^4 f_4$, 
$\bar{f}_5 = M^4 f_5$, $\bar{f}_6 = M^5 f_6$, $\bar{f}_{10} = M^3 f_{10}$, $\bar{f}_{11} = M^4 f_{11}$, 
$\bar{f}_{12} = M^5 f_{12}$.  
We have then used the expansion of the amplitudes $f_i$ given in (\ref{eq:fexpand}), 
with dimensionless quantities $\bar{f}_i = \bar{f}_{i,0} + \order(\omega^2)$ ($i=1,2,5,6,11,12$) and 
$\bar{f}_i = \bar{f}_{i,1} q\cdot p / M^2 + \order(\omega^3)$ ($i=4,10$). 
Note that six new phenomenological parameters are required at $1/M^4$ to match $c_{X7\dots X12}$. 
An expression for $c_{X4}$ in terms of scattering observables can be found using (\ref{eq:constrain4}).
The expressions (\ref{eq:virtual}) can be translated to a multitude of
other conventions for the observables of Compton scattering.%
\footnote{ 
For example, Compton scattering of real photons determines 
all coefficients, apart from $c_{X7}$ and $c_{X12}$,
in terms of the conventional electric and magnetic 
polarizabilities $\alpha_E, \beta_M$ 
and Ragusa's~\cite{Ragusa:1993rm} spin polarizabilities $\gamma_i$: 
$M^3 \alpha_E/\alpha = -\bar{f}_{1,0} - 4 \bar{f}_{2,0}$, 
$M^3 \beta_M/\alpha = \bar{f}_{1,0}$,
$M^4\gamma_1/\alpha = -8\bar{f}_{4,1} - 4\bar{f}_{10,1}-4\bar{f}_{11,0}$,
$M^4 \gamma_2/\alpha = 4 \bar{f}_{10,1}$,
$M^4 \gamma_3/\alpha = 4\bar{f}_{6,0} + 2\bar{f}_{11,0}$,
$M^4 \gamma_4/\alpha = -4\bar{f}_{10,1} - 2\bar{f}_{11,0}$,
where $\alpha$ is the fine structure constant. Compare with Eq.~(28) of~\cite{Drechsel:1998zm}.
}

\section{Pure photon and four-fermion operators \label{sec:fourfermion}}

So far our analysis has focused on the one-fermion sector.  We have derived the
form of the Lagrangian appropriate, e.g., to a proton
in a background electromagnetic field. 
Let us consider the complete QED theory including dynamical photon, as
well as a lepton (electron or muon) field.   
The case of a nonrelativistic lepton is appropriate to bound state hydrogen 
studies, or very low-energy lepton-nucleon (e.g. muon-proton) scattering, where $E \ll M_\chi, M$. 
We first consider this case, constructing the operator basis, deriving 
coefficient relations and identifying redundant operators. 
We then turn to a brief discussion of 
the case of a relativistic lepton, appropriate to e.g. low-energy electron-proton
scattering with $m_\ell, E \ll M$. 

\subsection{Pure photon operators}

The pure gauge sector for NRQED is the well-known Euler-Heisenberg Lagrangian. 
Enforcing parity and time-reversal symmetry and neglecting total derivatives 
we find 
\be\label{eq:Lphoton}
{\cal L}_\gamma = -\frac14 F_{\mu\nu} F^{\mu\nu} 
+ c_{V2} {F_{\mu\nu} [\partial^2 F^{\mu\nu} ] \over M^2} 
+ c_{V4} {F_{\mu\nu} [\partial^4 F^{\mu\nu} ] \over M^4} 
+ c_{E1} g^2 { (F_{\mu\nu} F^{\mu\nu} )^2 \over M^4} 
+ c_{E2} g^2 { F^{\mu}_{\,\,\nu} F^\nu_{\,\,\rho} F^\rho_{\,\,\sigma} F^\sigma_{\,\,\mu} \over M^4 } + \dots \, .
\ee
In the literature one finds also the operator basis $(\frac12\epsilon^{\mu\nu\rho\sigma} F_{\mu\nu} F_{\rho\sigma})^2=16(\bm E\cdot \bm B)^2$ and $(F_{\mu\nu}F^{\mu\nu})^2=4(\bm E^2-\bm B^2)^2$. The relation to the basis above is obtained via $4{ F^{\mu}_{\,\,\nu} F^\nu_{\,\,\rho} F^\rho_{\,\,\sigma} F^\sigma_{\,\,\mu}}=2(F_{\mu\nu}F^{\mu\nu})^2+(\frac12\epsilon^{\mu\nu\rho\sigma} F_{\mu\nu} F_{\rho\sigma} )^2$.
The coefficients $c_{V2}$ and $c_{V4}$ may be set to zero through field redefinitions 
on $A^\mu$, as discussed in Sec. \ref{sec:redundancies} below.

\subsection{Four-fermion operators\label{sec:4f}}

Consider four-fermion operators relevant for processes in the one-nucleon, 
one-lepton sector. 
We enforce Hermiticity and invariance under 
parity, time-reversal and rotational symmetries.  
We use the notation $\overleftarrow{D}$ for a 
covariant derivative acting to the left, $X\overleftarrow{D}^i \equiv [\partial^i X] + igZ X A^i$, 
and define $D_+ \equiv  D + \overleftarrow{D} $, $D_- \equiv D - \overleftarrow{D}$.  
Having performed field redefinitions to eliminate operators with
time derivatives acting on heavy fermions, 
the Lagrangian in this sector, through $1/M^4$, is
\begin{align}\label{eq:psichi}
{\cal L}_{\psi \chi} &= { d_1 \over M^2} \psi^\dagger \sigma^i \psi  \ \chi^\dagger \sigma^i \chi
+ {d_2 \over M^2} \psi^\dagger \psi \ \chi^\dagger \chi 
+ {d_3\over M^4} \psi^\dagger D_+^i \psi \ \chi^\dagger D_+^i \chi 
+ {d_4\over M^4} \psi^\dagger D_-^i \psi \ \chi^\dagger D_-^i \chi 
\nl
&\quad
+ {d_5\over M^4} \psi^\dagger (\bm{D}^2 + \overleftarrow{\bm{D}}^2 )\psi \ \chi^\dagger\chi 
+ {d_6\over M^4} \psi^\dagger\psi \ \chi^\dagger (\bm{D}^2 + \overleftarrow{\bm{D}}^2 )\chi
\nl
&\quad
+ {g d_7\over M^4} \psi^\dagger \bm{\sigma}\cdot\bm{B} \psi \ \chi^\dagger\chi 
+ {i d_8\over M^4} \epsilon^{ijk} \psi^\dagger \sigma^i D_-^j \psi \ \chi^\dagger D_+^k \chi 
+ {i d_9\over M^4} \epsilon^{ijk} \psi^\dagger \sigma^i D_+^j \psi \ \chi^\dagger D_-^k \chi 
\nl
&\quad
+ {g d_{10}\over M^4} \psi^\dagger\psi \ \chi^\dagger \bm{\sigma}\cdot\bm{B} \chi 
+ {i d_{11}\over M^4} \epsilon^{ijk} \psi^\dagger D_+^k \psi \ \chi^\dagger \sigma^i D_-^j \chi 
+ {i d_{12}\over M^4} \epsilon^{ijk} \psi^\dagger  D_-^k \psi \ \chi^\dagger \sigma^i  D_+^j \chi 
\nl
&\quad
+ {d_{13}\over M^4} \psi^\dagger \sigma^i D_+^j \psi \ \chi^\dagger \sigma^i D_+^j \chi 
+ {d_{14}\over M^4} \psi^\dagger \sigma^i D_-^j \psi \ \chi^\dagger \sigma^i{D}_-^j \chi 
+ {d_{15}\over M^4} \psi^\dagger \bm{\sigma}\cdot \bm{D}_+ \psi \ \chi^\dagger \bm{\sigma}\cdot \bm{D}_+ \chi 
\nl
&\quad
+ {d_{16}\over M^4} \psi^\dagger \bm{\sigma}\cdot \bm{D}_- \psi \ \chi^\dagger \bm{\sigma}\cdot \bm{D}_- \chi 
+ {d_{17}\over M^4} \psi^\dagger \sigma^i D_-^j \psi \ \chi^\dagger \sigma^j D_-^i \chi 
\nl
&\quad
+ {d_{18}\over M^4} \psi^\dagger \sigma^i ( \bm{D}^2 + \overleftarrow{\bm{D}}^2 ) \psi \ \chi^\dagger \sigma^i \chi
+ {d_{19}\over M^4} \psi^\dagger \sigma^i ( D^i D^j + \overleftarrow{D}^j \overleftarrow{D}^i) \psi \ \chi^\dagger \sigma^j \chi 
\nl
&\quad
+ {d_{20}\over M^4} \psi^\dagger \sigma^i \psi \ \chi^\dagger \sigma^i ( \bm{D}^2 + \overleftarrow{\bm{D}}^2 ) \chi
+ {d_{21}\over M^4} \psi^\dagger \sigma^i \psi \ \chi^\dagger \sigma^j ( D^i D^j + \overleftarrow{D}^j \overleftarrow{D}^i) \chi  
\,.
\end{align} 
Here $\chi$ is the nonrelativistic lepton field with mass $M_\chi$ and for notational 
simplicity we write all operators in terms of the common mass scale $M$.%
\footnote{
Note that the coefficients $d_{1,2}$ in (\ref{eq:psichi}) 
are related to those of Caswell and Lepage~\cite{Caswell:1985ui}
by a factor $M_\chi/M$. 
}
Covariant derivatives appearing within a fermion bilinear in (\ref{eq:psichi}) are understood 
to act only on fields in that bilinear. 
The heavy field $\psi$ transforms under boosts as in (\ref{eq:lorentz}). 
Recalling that $\bm{q}$ in (\ref{eq:lorentz}) is related to the mass-independent 
infinitesimal boost parameter by $\bm{\eta} = -\bm{q}/M$, 
the transformation law for $\chi$ is obtained by the replacement 
$M \to r M$ and $q \to rq$, where we define $r\equiv M_\chi/M$. 
We thus find 
\begin{align}\label{eq:L4f}
&\delta {\cal L}_{\psi\chi} 
= \frac{1}{M^4} \bigg\{ 
\psi^\dagger i \bm{q} \cdot \bm{D}_- \psi \ \chi^\dagger \chi \left[ \frac{d_2}{2} - 2rd_4 - 2d_5 \right]
+ \psi^\dagger \psi \ \chi^\dagger i \bm{q} \cdot \bm{D}_- \chi \left[ \frac{d_2}{2r} - 2d_4 - 2rd_6 \right] \nl
&
+\psi^\dagger \bm{\sigma} \cdot \bm{q} \times \bm{D}_+ \psi  \ \chi^\dagger \chi \left[  -\frac{d_2}{4}  + \frac{d_1}{4r} - 2d_8 - 2rd_9 \right] 
+ \psi^\dagger i \bm{q} \cdot \bm{D}_- \sigma^i \psi \ \chi^\dagger \sigma^i \chi \left[ \frac{d_1}{2}  - 2rd_{14} - 2d_{18} \right] 
\nl
&
+\psi^\dagger \psi \chi^\dagger \bm{\sigma} \cdot \bm{q} \times \bm{D}_+ \chi \left[  -\frac{d_2}{4r} + \frac{d_1}{4} - {2rd_{11}} - 2d_{12} \right] 
+ \psi^\dagger \sigma^i \psi \chi^\dagger i \bm{q} \cdot \bm{D}_- \sigma^i \chi  \left[ \frac{d_1}{2r} - 2d_{14} - 2rd_{20}  \right]
\nl
&
+ \psi^\dagger i \bm{\sigma} \cdot \bm{D}_- \psi \ \chi^\dagger \bm{\sigma} \cdot \bm{q} \chi \left[ \frac{d_1}{4} - {2rd_{16}} - d_{19} \right] 
+ \psi^\dagger \bm{\sigma} \cdot \bm{q}  \psi \ \chi^\dagger i \bm{\sigma} \cdot \bm{D}_- \chi \left[ \frac{d_1}{4r} - 2d_{16} - rd_{21} \right] 
\nl
&
+\psi^\dagger i \bm{\sigma} \cdot \bm{q} D_-^i \psi \ \chi^\dagger \sigma^i \chi \left[ -\frac{d_1}{4} - 2rd_{17} - d_{19} \right]
+\psi^\dagger \sigma^i \psi  \chi^\dagger i \bm{\sigma} \cdot \bm{q} D_-^i \chi  \left[ -\frac{d_1}{4r} - 2d_{17} -rd_{21} \right]
\bigg\}  
\nl
& 
\quad + \order(1/M^5)  \,.
\end{align}
This enforces the relations
\begin{align}\label{eq:fourrelations}
  rd_4 +d_5  &= {d_2\over 4} \, , \quad  
d_5  = r^2 d_6 \,, \quad
8r(d_8 + rd_9) = -rd_2 + d_1  \,, \quad 
8r ( rd_{11}+d_{12} ) = -d_2  + rd_1 \,, 
\nl
rd_{14} + d_{18} &= {d_1\over 4} \, , \quad  
d_{18}  = r^2 d_{20}  \,, \quad 
2rd_{16} + d_{19} = {d_1\over 4} \,, \quad 
r(d_{16}+d_{17})+d_{19}=0 \,, \quad
d_{19} = r^2 d_{21},
\end{align}
implying a total of twelve independent four-fermion operators through 
$1/M^4$, including two at order $1/M^2$.   
By performing field redefinitions on the gauge field $A^\mu$, 
some of these four-fermion operators are found to mix with one-heavy 
particle sector operators, as discussed in Section \ref{sec:redundancies} below.

The Lagrangian (\ref{eq:psichi}), with constraints (\ref{eq:fourrelations}), applies
to the case of distinct heavy particles represented by $\psi$, $\chi$, with 
arbitrary mass ratio $M_\chi/M$. 
For certain applications, e.g., positronium or heavy quarkonium bound states, 
the fields $\psi$ and $\chi$ can be taken to 
represent particle-antiparticle pairs with $r=M_\chi/M=1$.   
Charge conjugation symmetry is then implemented by enforcing 
invariance under $\psi \leftrightarrow \chi$, thus reducing the basis of operators.   
This case has been investigated for QCD through $\order(1/M^4)$ by Brambilla 
{\it et al.}~\cite{Brambilla:2008zg}.   
We find that our basis of four-fermion operators (\ref{eq:psichi}) and constraints (\ref{eq:fourrelations}) 
are equivalent to those found in Ref.~\cite{Brambilla:2008zg} for this special case.%
\footnote{The difference between Abelian and non-Abelian gauge fields 
is trivial for four-fermion operators
through this order.}

\subsection{Field redefinitions and redundant operators} \label{sec:redundancies} 

With a dynamical photon field, we may 
perform field redefinitions that maintain 
reality and gauge, parity, time-reversal and rotational symmetries.
In order to avoid upsetting the previously determined coefficient relations, 
we must also maintain the transformation law for $A^\mu$ as a four-vector under Lorentz transformations, i.e., 
\be\label{eq:Atransform}
A^0 \to A^0 - {1\over M} \bm{q}\cdot\bm{A} \,, \quad
\bm{A} \to \bm{A} - {1\over M} \bm{q} A^0 \,. 
\ee
Let us write 
\be\label{eq:DA}
A_\mu = A^{\prime}_\mu + \Delta_\gamma A_\mu + \Delta_{\psi} A_\mu + \Delta_\chi A_\mu + \dots \,. 
\ee
For the pure gauge field terms the most general expression is  
\be \label{eq:vacpol}
\Delta_\gamma A^\mu = a_{\gamma 1} {\partial_\nu
F^{\nu\mu} \over  M^2}  + a_{\gamma 2} { \partial^2 \partial_\nu F^{\nu\mu}
\over M^4} + \order(1/M^6) \,. 
\ee
Terms involving the heavy fermion $\psi$ take the form 
\begin{align}\label{eq:psi_A}
{\Delta_\psi A^\mu\over g} &= 
\tilde{a}_{\psi 1} { \overline{\Psi}_v \gamma^\mu \Psi_v \over M^2 }
+ \tilde{a}_{\psi 2} {\partial_\alpha (\overline{\Psi}_v \sigma^{\alpha\mu} \Psi_v) \over M^3}
+ \tilde{a}_{\psi 3}g {\overline{\Psi}_v \{ \gamma^\mu , \, \sigma^{\alpha\beta} F_{\alpha\beta} \} \Psi_v \over M^4 } 
+ \tilde{a}_{\psi 4} { \partial^2 (\overline{\Psi}_v \gamma^\mu \Psi_v) \over M^4 }   
\nl
&\quad 
+ \tilde{a}_{\psi 5}g { \overline{\Psi}_v \sigma^{\mu\alpha} \{ {\cal V}^\beta , F_{\alpha\beta} \} \Psi_v \over M^4} 
+ \order( 1/M^5 ) \,,
\end{align}
where we have employed the invariant operator formalism of Sec.~\ref{sec:invariant}.  
In particular, $\Psi_v = \Gamma \psi_v$ with $\Gamma$ from (\ref{eq:Gamma}) 
and $\psi_v$ from (\ref{eq:psi}), 
expressed in terms of the field $\psi^\prime_v \equiv \psi$  
with canonical Lagrangian (\ref{eq:abelian}).  
As an alternative to the invariant operator formalism employed in (\ref{eq:psi_A}) 
we may expand $\Delta_\psi A^0$ and $\Delta_\psi \bm{A}$ in a series of rotationally
invariant operators with arbitrary coefficients, and subsequently constrain these coefficients
using (\ref{eq:Atransform}).   The result is equivalent to (\ref{eq:psi_A}), with five
free parameters through $\order(1/M^4)$, 
\begin{align}\label{eq:DApsi}
{\Delta_\psi A^0 \over g} &=  
a_{\psi 1} {\psi^\dagger \psi \over M^2}  
+ a_{\psi 2} { \bm{\partial}^2(\psi^\dagger \psi) \over M^4 } 
-i\!\left({a_{\psi 1}\over 4} - a_{\psi 4} \right)\! { \psi^\dagger  \bm{\sigma}\cdot \overleftarrow{\bm{D}} \times \bm{D} \psi \over M^4 } 
+ a_{\psi 3}g {\psi^\dagger \bm{\sigma}\cdot\bm{B} \psi \over M^4} 
 + \order(1/M^5) , 
\nl
{\Delta_\psi \bm{A}\over g} &=
-{a_{\psi 1}} { \psi^\dagger i \bm{D}_- \psi \over 2 M^3} 
+ a_{\psi 4} { \bm{\partial}\times (\psi^\dagger\bm{\sigma} \psi) \over M^3 } 
+ a_{\psi 5} g { \psi^\dagger \bm{\sigma}\times \bm{E} \psi \over M^4 } 
+ \order(1/M^5) 
\,. 
\end{align}
The expansion of $\Delta_\chi A^\mu$ is obtained from (\ref{eq:DApsi}) 
with the replacements $\psi\to\chi$, $M\to M_\chi$, $Z\to Z_\chi$ and $a_{\psi\,i} \to a_{\chi\,i}$. 
In terms of the field $A_\mu^\prime$ in (\ref{eq:DA}), 
we find in 
the pure photon sector, 
\be\label{eq:photonshift}
\delta c_{V2} = -\frac12 a_{\gamma 1} \,, \quad 
\delta c_{V4} = -\frac12 a_{\gamma 2} -\frac14 a_{\gamma 1}^2 + 2 a_{\gamma 1} c_{V2} \,,
\ee
while for the $\psi$ sector, 
\begin{align}\label{eq:psishift}
\delta c_{D} &= -8 Z a_{\gamma 1} + 8 a_{\psi 1}  \,, \quad 
\delta c_{W1} = -4c_F a_{\gamma 1} + 8 a_{\psi 4} \,, \quad 
\delta c_{A2} = -16 Z^2 a_{\gamma 1} + 16 Z a_{\psi 1} \,, \quad
\nl
\delta c_{X3} &= -{c_D a_{\gamma 1}\over 8} + Z a_{\gamma 2} - a_{\gamma 1} a_{\psi 1} 
+ 4 c_{V2} a_{\psi 1} + a_{\psi 2} 
\,, \quad
\delta c_{X7} = -{c_S Z a_{\gamma 1} \over 4} + a_{\psi 3}\,, \quad
\nl
\delta c_{X8} &= c_F Z a_{\gamma 1} - {c_F a_{\psi 1}\over 2} - Z a_{\psi 4}  \,, \quad
\delta c_{X9} = -{c_F^2 a_{\gamma 1} \over 2} + c_F a_{\psi 4} \,, \quad
\delta c_{X11} = {c_F^2 a_{\gamma 1} \over 2} - c_F a_{\psi 4} \,, 
\nl
\delta c_{X12} &= {c_S Z a_{\gamma 1} \over 2} + a_{\psi 5}  \,.
\end{align}
Similar relations hold for the Wilson coefficients 
$c_{i}^{(\chi)}$ in the $\chi$ Lagrangian, defined as in (\ref{eq:abelian}), 
with $\psi\to \chi$, $Z\to Z_\chi$, $M\to M_\chi$, $c_i \to c_i^{(\chi)}$. 
Finally, for the four-fermion operator coefficients, 
\begin{align}\label{eq:fourshift}
{\delta d_2\over g^2} &= -Z_\chi a_{\psi 1} - {Z a_{\chi 1} \over r^2} \,, \quad
{\delta d_3 \over g^2} = {c_D^{(\chi)}a_{\psi 1} \over 8 r^2} + {c_D a_{\chi 1} \over 8 r^2} + Z_\chi a_{\psi 2} + {Z a_{\chi 2}\over r^4} \,,
\nl
{\delta d_4 \over g^2} &= -{Z_\chi a_{\psi 1}\over 4 r} - {Z a_{\chi 1} \over 4 r^3} \,, \quad
{\delta d_7 \over g^2} = -{Z Z_\chi \over 4} \left(a_{\psi 1} - 4 a_{\psi 4} \right) - Z_\chi a_{\psi 3} \,,
\nl
{\delta d_8 \over g^2} &= {Z_\chi \over 8} \left( a_{\psi 1} - 4 a_{\psi 4} \right) - {c_S a_{\chi 1} \over 8 r^2} \,, \quad
{\delta d_{10} \over g^2} = -{Z Z_\chi \over 4 r^4} \left( a_{\chi 1} - 4 a_{\chi 4} \right) - {Z a_{\chi 3}\over r^4} \,, 
\nl
{\delta d_{11} \over g^2} &= - {c_S^{(\chi)} a_{\psi 1} \over 8 r^2} + {Z \over 8 r^4} \left( a_{\chi 1} - 4 a_{\chi 4} \right) \,, \quad
{\delta d_{13} \over g^2} = -{\delta d_{15} \over g^2} = {c_F^{(\chi)} a_{\psi 4} \over 2 r} + {c_F a_{\chi 4} \over 2 r^3} \,.
\end{align} 
The coefficient relations (\ref{eq:constrain12}), (\ref{eq:constrain4}) and (\ref{eq:fourrelations}) 
are preserved, since by construction the Lorentz transformation
properties of $A^\mu$ are unchanged and hence the boost transformation
rules (\ref{eq:lorentz}) and (\ref{eq:lorentz2}) still apply.

We may use (\ref{eq:photonshift}) 
to eliminate vacuum polarization terms $c_{V2}$ and $c_{V4}$ in favor of compensating 
terms in (\ref{eq:psishift}).  
Similarly, (\ref{eq:psishift}), together with the analogous relations for $c_{i}^{(\chi)}$, and 
(\ref{eq:fourshift}), can be used to eliminate 10 linear combinations of 
Wilson coefficients for two-fermion and four-fermion operators. 
Different applications may favor elimination of different operators.%
\footnote{
We have not specified gauge fixing and source terms, which are also affected by field redefinitions.
}

\subsection{Relativistic lepton \label{sec:rel}}

For applications such as lepton-nucleon scattering at energies $m_\ell, E \ll M$ 
(e.g., low-energy electron-proton scattering), the relevant effective theory involves 
a heavy fermion $\psi$ with mass $M$ (e.g., the proton) interacting with an 
electromagnetically charged relativistic fermion $\ell$ with mass $m_\ell$ 
(e.g., the electron). Let us briefly discuss this case. 
Enforcing parity, time-reversal, gauge, Lorentz, as well as chiral symmetry at $m_\ell=0$, 
we find the leptonic interactions with the photon, 
\be
{\cal L}_{\ell} =  \bar{\ell}\left[ i\Dslash - m_\ell + g c_F^{(\ell)} m_\ell {\sigma^{\mu\nu} F_{\mu\nu}  \over M^2} 
+ g c_2^{(\ell)} m_\ell  {D^2\over M^2} + g c_D^{(\ell)} {  [\partial^\mu F_{\mu\nu} ]  \gamma^\nu\over M^2} + \order(1/M^4) \right] \ell \,,
\ee
where we assume field redefinitions have been performed to remove power suppressed terms involving 
$(i\Dslash-m_\ell)\ell$.  

Having performed field redefinitions to eliminate operators with 
time derivatives acting on fermion fields, 
the Lagrangian for the nucleon-relativistic lepton sector through ${\cal O} (1/M^3)$ is
\begin{align}\label{eq:bi}
{\cal L}_{\psi \ell} &= 
{b_1\over M^2} \psi^\dagger  \psi  \ \bar{\ell} \gamma^0 \ell  
+ {b_2 \over M^2} \psi^\dagger \sigma^i \psi \ \bar{\ell} \gamma^i \gamma_5 \ell  
+ {b_3\over M^3} \psi^\dagger \psi \ m_\ell \bar{\ell} \ell
+ {b_4\over M^3} \psi^\dagger i D_-^i \psi \ \bar{\ell} \gamma^i \ell
\nl
&\quad
+ {b_5\over M^3} \psi^\dagger \psi \bar{\ell} i\bm{\gamma}\cdot\bm{D}_- \ell 
+ {b_6\over M^3} \epsilon^{ijk} \psi^\dagger \sigma^i \psi \ m_\ell \bar{\ell} \sigma^{jk} \ell 
+ {b_7\over M^3} \epsilon^{ijk} \psi^\dagger \sigma^i  \psi   \ \bar{\ell} \gamma^j D_+^k \ell
\nl
&\quad
+ {b_{8}\over M^3} \psi^\dagger \sigma^i \psi \ \bar{\ell} \gamma^0 \gamma_5 i D_-^i \ell
+ {b_{9}\over M^3} \psi^\dagger \sigma^i iD_-^i \psi \ \bar{\ell} \gamma^0 \gamma_5 \ell
+ \order(1/M^4) 
,
\end{align} 
where $\ell$ is the relativistic lepton field with mass  $m_\ell$ 
and $\sigma^{ij} \equiv \frac{i}{2} [ \gamma^i, \gamma^j]$.  
The heavy field $\psi$ 
transforms under boosts as in (\ref{eq:lorentz}), 
while $\ell$ transforms under finite-dimensional representations of the Lorentz group 
in the usual way. Under Lorentz transformation, we thus find
\be
\delta {\cal L}_{\psi \ell} = - \frac{1}{M^3} \psi^\dagger \psi \bar{\ell} \bm{\gamma} \cdot \bm{q} \ell \left( b_1 + 2 b_4 \right) 
- \frac{1}{M^3} \psi^\dagger \bm{\sigma} \cdot \bm{q} \psi \bar{\ell} \gamma^0 \gamma_5 \ell \left( b_2 +2b_9 \right) + {\cal O}(1/M^4).
\ee
This enforces the relations
\be\label{eq:brel}
b_4 = -\frac12 b_1 \,, \quad b_9 = -\frac12 b_2 \,,
\ee
leaving seven operators in this sector through order $1/M^3$, including two at order $1/M^2$.

By performing field redefinitions on the gauge field $A^\mu$, some of these 
four-fermion operators are found to mix with one-heavy particle operators. 
In addition to the contributions $\Delta_\gamma A^\mu$ and $\Delta_\psi A^\mu$ 
from (\ref{eq:DA}) we may employ 
\be
\Delta_\ell A^\mu = g a_{\ell 1} { \bar{\ell} \gamma^\mu \ell \over M^2} + \order(1/M^4) \,.
\ee
We thus find the modified couplings in ${\cal L}_\ell$, 
\begin{align}
\delta c_D^{(\ell)} &= - Z_\ell a_{\gamma 1} + a_{\ell 1} \,, 
\end{align}
and for the four-fermion operators in ${\cal L}_{\psi\ell}$, 
\begin{align}
{\delta b_1\over g^2} &= -Z a_{\ell 1} - Z_\ell a_{\psi 1} \,, \quad
{\delta b_7 \over g^2} = -Z_\ell a_{\psi 4} -  \frac12 c_F a_{\ell 1} \,,   
\end{align}
with relation (\ref{eq:brel}) remaining intact. 

\section{Applications \label{sec:apps}} 

Applications of the NRQED Lagrangian for the nucleon 
include the computation 
of proton structure effects in atomic bound states,  
the model-independent analysis of radiative corrections
to low-energy  lepton-nucleon scattering, 
and the study of static electromagnetic properties of
nucleons.  Let us discuss sample applications in each of these areas. 

\subsection{Proton structure in atomic bound states\label{sec:bound}} 

As a sample computation involving the $1/M^4$ NRQED Lagrangian,%
\footnote{
For bound state applications it may be computationally efficient to further integrate out 
off shell momentum modes and/or higher Fock states 
to arrive at an explicitly $v/c$ expanded NRQED~\cite{Caswell:1985ui,Kinoshita:1995mt}, 
or fixed particle number quantum mechanics~\cite{Hill:2000qi}.  
Our purpose here is to examine the impact of nucleon structure, 
as described by shifts in the Wilson coefficients of the theory described 
by (\ref{eq:abelian}), (\ref{eq:Lphoton}), (\ref{eq:psichi}).
}
let us analyze the effects of nuclear structure on the two-photon exchange 
contribution to the $2S-2P$ Lamb shift in hydrogenic bound states; this
contribution is the subject of intense 
scrutiny~\cite{Pachucki:1999zza,Pineda:2004mx,Miller:2011yw,Carlson:2011zd,Hill:2011wy,Birse:2012eb,WalkerLoud:2012bg} 
due to the discrepant measurements in muonic and electronic hydrogen~\cite{Pohl:2010zza}. 
The dominant theoretical uncertainties in the muonic hydrogen Lamb shift arise from 
proton structure, represented by contributions through $\order(\alpha)$ 
to $c_D$ in (\ref{eq:abelian}), arising from first order vertex corrections, 
and $\order(\alpha^2)$ contributions to $d_2$ in (\ref{eq:psichi}) arising from 
two-photon exchange~\cite{Hill:2011wy}.  
We let $\psi$ of Sec.~\ref{sec:abelian} denote the proton of mass $M$, and 
$\chi$ of Sec.~\ref{sec:fourfermion} denote the electron (or muon) of mass $M_\chi=m_e$. 
Electric charge assignments are given after (\ref{eq:abelian}). 
We focus here on the model-independent result for the leading 
two-photon exchange corrections in the limit $m_e/M \ll 1$. 

Structure-dependent corrections to $d_2$ lead to first order energy shifts 
in hydrogen,%
\footnote{
As mentioned in Sec.~\ref{sec:4f}, we have used powers of $1/M$ (not $1/M_\chi$) 
to define four-fermion Wilson coefficients in (\ref{eq:psichi}).  
We thus have $d_{1,2}/M^2 = d^{\rm CL}_{1,2}/({m_e M})$ where $d_{2}^{\rm CL}$ 
is the coefficient of Caswell and Lepage~\cite{Caswell:1985ui}, 
also employed in \cite{Hill:2011wy}. 
}
\be\label{eq:Enl}
\delta E(n,\ell) = \delta_{\ell 0} {m_r^3 \alpha^3 \over \pi n^3} \left( 
- { \delta d_2 \over M^2}  \right) \,,
\ee
where $(n,\ell)$ are principal and orbital quantum numbers
and 
$m_r = m_e M/(m_e + M)$ is the reduced mass.
The matching condition for $d_2$ involves a weighted 
moment of the structure functions for forward Compton scattering on the proton.
From \cite{Hill:2011wy}, the necessary hadronic input is contained in 
\be\label{eq:d2match}
{\delta d_2 \over M^2}  
= 
-{4 m_e \alpha^2 \over \pi M}
\int_{-1}^1 dx   \sqrt{1-x^2} 
\int_0^\infty {dQ\over Q} 
{\left[ (1+2x^2)W_1 - (1-x^2) m_p^2 W_2 \right]
\over Q^2 + 4 m_e^2 x^2 } + {\rm subtractions} \,, 
\ee
where the subtractions depend on infrared and ultraviolet regulators.  
The structure functions in this expression are evaluated at $\nu^2 = - 4 x^2 M^2 Q^2$. 
They are defined by the forward proton matrix element,
\be\label{eq:comptongeneral}
W^{\mu\nu}(k,q,s) \equiv 
i \int d^4x\, e^{iq\cdot x} \langle \bm{k},s| T\{ J_{\rm e.m.}^\mu(x) J_{\rm e.m.}^\nu(0) \} | \bm{k},s \rangle \,. 
\ee
For the present application we require the spin-averaged component symmetric in $\mu,\nu$, 
\begin{align}\label{eq:forwardcompton}
W_S^{\mu\nu} &= \frac12 \sum_s W^{\mu\nu} 
= 
\left( - g^{\mu\nu} + {q^\mu q^\nu \over q^2} \right) W_1(\nu,Q^2)
+ 
\left( k^\mu - {k\cdot q \,q^\mu  \over q^2} \right) 
\left( k^\nu - {k\cdot q \, q^\nu  \over q^2} \right) W_2(\nu,Q^2) \,,
\end{align} 
where $Q^2=-q^2$ and $\nu=2 k\cdot q$.  
Our normalizations are such that 
for a point particle, $W_1=2\nu^2/(Q^4-\nu^2)$ and $W_2=8Q^2/(Q^4-\nu^2)$.  

Let us proceed to evaluate the finite shift $\delta d_2$ due to proton structure,
and hence $\delta E$ in (\ref{eq:Enl}),
in the limit $m_e \ll M$.  Details of this computation are presented elsewhere. 
The essential input from NRQED is the evaluation of 
$W_i(0,Q^2)$ through $\order(Q^2)$, 
\begin{align}\label{eq:Wi0}
W_1(0,Q^2) &= -2 + 2c_F^2 + {Q^2\over 2 M^2} \left( c_F^2 - 2 c_F c_{W1} + 2 c_M + c_{A1} \right) + \order(Q^4)  \,,
\nl
M^2 W_2(0,Q^2) &= {8 M^2\over Q^2} + 2 c_F^2 - 2 c_D 
+ {Q^2 \over 2 M^2} \bigg[ -1 + c_F^2 - c_D +\frac14 c_D^2 + 2 c_M - 2 c_F c_{W1} - \frac12 c_{A2} 
\nl &\quad 
+ 32(c_{X1}+c_{X2}+c_{X3} ) \bigg] + \order(Q^4) \,,
\end{align} 
and the evaluation of $W_1(\nu,0)$ through $\order(\nu^2)$, 
\begin{align}
W_1(\nu,0) &= -2 + {\nu^2\over 8 M^4} \left( -c_F^2 + c_F c_S + 2 c_M - \frac12 c_{A2} \right) 
+ \order(\nu^4) \,,
\nl
 \lim_{Q^2\rightarrow0}W_2(\nu,Q^2) &= 0 \,, 
\end{align}
where the final relation follows from requiring that the 
Compton amplitude  (\ref{eq:forwardcompton}) 
is finite at $Q^2 \rightarrow 0$. 
Note in particular that $W_2(0,Q^2)$ 
relies on $\order(1/M^4)$ one-photon vertices involving $c_{X1}$, $c_{X2}$, $c_{X3}$. 
This occurs because the $\order(Q^2)$ terms in $W_2(0,Q^2)$ arise at third order in the $Q^2/M^2$ 
expansion. 
Employing these results in the matching relation (\ref{eq:d2match}) 
yields the leading contribution to $S$-state energies at $m_e/M\to 0$, 
\begin{multline}\label{eq:logenergy}
\delta E(nS) = {m_e^3 \alpha^5 \over \pi n^3} {m_e \over M^3} \Big\{ 
\log {m_e \over M} \big[ 
{M^3} \left( 5 \alpha_E - \beta_M \right)/\alpha 
- 3 a_p(1+a_p) + 2 M^2 (r_E^p)^2 
\big] 
+ \dots 
\Big\} \,. 
\end{multline}
Our result for the coefficient of the logarithm containing $\alpha_E$ and $\beta_M$ 
was obtained in a temporal gauge analysis in \cite{Khriplovich:1997fi}.    
We have here used NRQED to establish the complete logarithmically enhanced contribution. 
The expansion at small lepton mass is not possible for muonic hydrogen owing to
low hadronic scales scales $m_\mu \sim m_\pi \sim m_\Delta-m_p$.   
Further analysis for this case will be presented elsewhere.

\subsection{Radiative corrections to lepton-nucleon scattering\label{sec:scattering}}

\begin{figure}[htb]
\begin{center}
\includegraphics[scale=0.8]{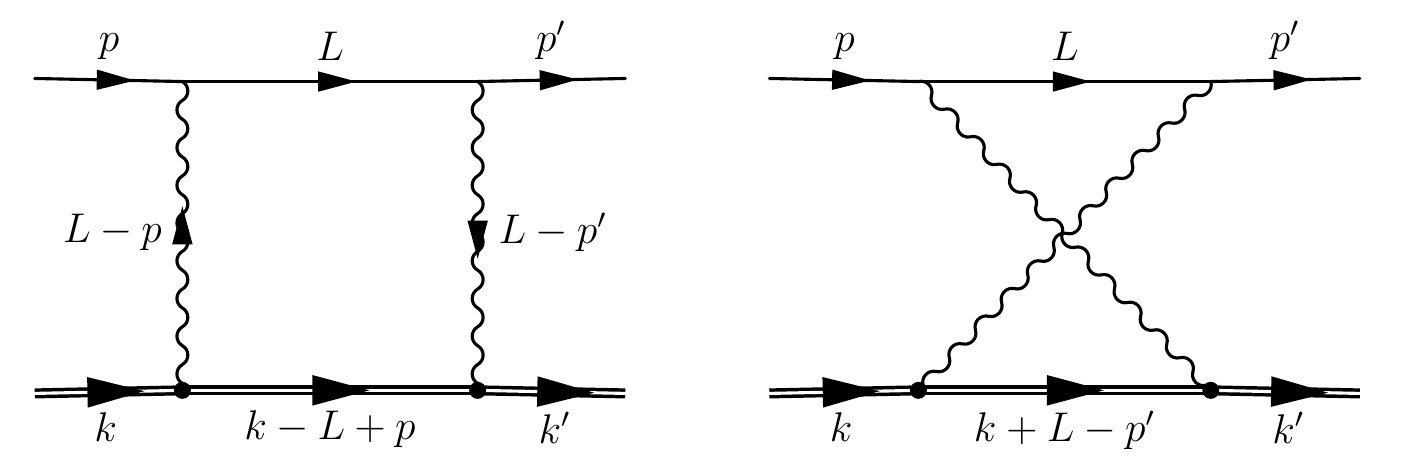}
\end{center}
\vspace{-5mm}
\caption{\label{fig:ep_scattering}
Diagrams contributing at leading order in $1/M$ to two-photon exchange 
in electron-proton scattering. 
}
\end{figure}

Extraction of key hadronic quantities from elastic lepton-nucleon scattering, 
such as the charge and magnetic radii of the proton, 
demands rigorous control over radiative corrections. 
Traditional analyses of the two-photon exchange contribution to elastic electron-proton scattering
resort to hadronic models involving phenomenological form factor 
insertions into point-particle Feynman diagrams~\cite{Mo:1968cg,Maximon:2000hm,Blunden:2003sp,Blunden:2005ew}, 
or require modeling of soft functions to evaluate expressions obtained in a hard-scattering
factorization framework~\cite{Chen:2004tw}. 
At low energies, $E\ll M$, a systematic and model-independent approach is provided by NRQED.  
This region has overlap with current and planned electron-proton and muon-proton scattering
measurements, and provides a rigorous test of hadronic models employed at higher energy. 

Consider the two-photon exchange corrections to low-energy electron-proton scattering
in the limit $m_\ell = m_e \ll E \ll M$, where
$E$ is the energy of the incident electron in the rest frame of the initial-state proton.  
The appropriate effective theory consists of a relativistic lepton and heavy proton, 
as described in Sec.~\ref{sec:rel}. 
The leading order NRQED diagrams, shown in Fig.~\ref{fig:ep_scattering}, are not sensitive to proton structure, yielding, 
in Feynman gauge, [electric charge assignments are given after (\ref{eq:abelian})]
\begin{align}
&i{\cal M} \approx e^4 \bar{u}(p^\prime) \gamma^0 \gamma^\mu \gamma^0 u(p)
\nl
& \,\, 
\times \int {d^4 L \over (2\pi)^4} 
{L_\mu \over L^2+ i0 } {1 \over (L-p)^2- \lambda^2+ i0} {1\over (L-p^\prime)^2 - \lambda^2+i0} 
\left( {1\over - L^0 + E+i0}  + {1\over L^0 - E+i0} \right) \,,
\end{align} 
where we have neglected the electron mass and 
used that $p^0-p^{\prime 0}$, $k^0-M$, and $k^{\prime 0} -M$ are subleading in the $1/M$ expansion to simplify the integrand.  A photon mass $\lambda$ is used to regulate infrared divergences. 
Evaluating the integral in the limit $\lambda \ll E$, 
we recover the expression for relativistic lepton scattering in an external Coulomb 
field~\cite{McKinley:1948zz,Dalitz:1951ah},%
\footnote{
There is an apparent typographical error in Eq.~(2.6) of \cite{Dalitz:1951ah} where in 
the second line the integral $I$ should have the opposite overall sign.  This affects only
the imaginary part of the $\order(\alpha)$ correction to the leading amplitude and 
is thus not relevant to first order radiative corrections in the cross section. 
}
\begin{align}
{\cal M} &= 
{4\pi \alpha \over Q^2} {\bar{u}(p) \gamma^0 u(p)} 
\bigg\{ 
1  + {\alpha} 
\bigg[ {\pi\over 2} {Q\over 2E + Q} + i \bigg( -2\log{\lambda \over Q} + {Q^2\over (2 E)^2 - Q^2} \log{Q\over 2E} \bigg) 
\bigg]
\nl
&\quad
+ \order[ \alpha^2, \lambda/E, m_e/E, E/M] \bigg\} \,. 
\end{align}
The power of the effective theory lies in the possibility to systematically 
compute corrections in powers of $E/M$ and to relate observables such as scattering
amplitudes and bound state energies, in a model-independent fashion. 
Sensitivity to the Wilson coefficient $c_F$ appears at order $1/M$ in the 
proton spin-dependent amplitude. 
At order $1/M^2$ there is a dependence on $c_D$, and on 
the constants $b_1$ and $b_2$ from (\ref{eq:bi}), 
which encode information on proton excitations and may be related to 
moments of the forward Compton amplitude.  

A similar analysis may be performed to compute radiative corrections to 
muon-proton scattering in the limit $E \ll  m_\mu \sim M$, 
where now the effective theory consists of heavy muon and proton fields, as
discussed in Sec.~\ref{sec:4f}.  Proton structure not captured by elastic 
form factors is first encountered at $\order(1/M^2)$, encoded in coefficients 
$d_1$ and $d_2$ in (\ref{eq:psichi}). 
NRQED may similarly be used to compute radiative backgrounds to 
searches for possible new low-mass and weakly coupled particles 
using low-energy electron-proton scattering~\cite{Bjorken:2009mm}.

\subsection{Nucleon polarizabilities and static properties of nucleons\label{sec:static}}

Having constructed the NRQED Lagrangian, it is straightforward to 
relate parameters extracted from scattering measurements to those 
determined by static nucleon properties.   Consider for simplicity the 
case of a neutral particle such as the neutron.%
\footnote{
Static properties of a charged particle require care in definition, see, e.g., \cite{Tiburzi:2008ma}. 
}
The shift in energy due to a constant external electromagnetic field 
is determined by the zero-momentum limit of the amputated 
two-point function, 
\begin{align}
- \delta M(\bm{E},\bm{B}) &= -\lim_{\bm{p}\to 0} \Sigma(\bm{p}) 
= c_F g { \bm{\sigma}\cdot \bm{B} \over 2M} 
+ c_{A1} g^2{ \bm{B}^2 - \bm{E}^2 \over 8 M^3}  
- c_{A2} g^2{ \bm{E}^2 \over 16 M^3 } + \ldots ,
\end{align}
and we may immediately identify the various static electromagnetic moments. 
The identification of $c_{i}$ as coefficients of a well-defined effective theory 
is essential for correctly relating static coefficients, 
as measured by lattice~\cite{Fiebig:1988en,Christensen:2004ca,Detmold:2006vu,Detmold:2010ts} 
or hadronic models, to scattering observables~\cite{Schumacher:2005an,Griesshammer:2012we}. 
For example, conventional definitions of the scalar electric ($\alpha_E$)  and magnetic ($\beta_M$) 
polarizabilities as determined by Compton scattering~\cite{Beringer:1900zz} require
that coefficients $c_{A1}$ and $c_{A2}$ are not proportional to these quantities, but
rather\footnote{Many approaches in the literature lack a systematic treatment of these 
effects~\cite{Pineda:2004mx,Schumacher:2005an,Detmold:2006vu,Detmold:2010ts,Griesshammer:2012we}.
The NRQED Lagrangian, constrained by Lorentz invariance, trivializes relations such as (\ref{eq:ebpols}) between
static nucleon properties and scattering observables.}
\begin{align} \label{eq:ebpols}
{4 M^3\over \alpha} \alpha_E &= -c_{A1} -\frac12 c_{A2} + Z^2 + 2 Z c_M + c_F c_S - c_F^2 \,,
\nl
{4 M^3\over \alpha} \beta_M &= c_{A1} - Z^2 \,. 
\end{align} 
Similar relations relate static spin polarizabilities to scattering observables: 
see (\ref{eq:virtual}) and the associated footnote. 

The coefficients of spin-dependent operators $c_{Xi}$ contribute to 
the spin-dependent structure functions of forward Compton scattering, obtained from the 
antisymmetric component of (\ref{eq:comptongeneral}), 
\begin{multline}
W^{\mu\nu}_A = \bar{u}(k) \bigg\{ H_1 \Big( [\gamma^\mu,\slash{q}]k^{\nu} -[\gamma^\nu, \slash{q}]k^{\mu} 
- [\gamma^\mu,\gamma^\nu]k\cdot q \Big) 
\\
+ H_2 \Big( [\gamma^\mu,\slash{q}] q^\nu -[ \gamma^\nu,\slash{q}] q^\mu - [\gamma^\mu,\gamma^\nu]q^2 \Big)
\bigg\} u(k) \,,
\end{multline}
where our normalization conventions are such that for a point particle $MH_1= Q^2/(Q^4-\nu^2)$, $H_2=0$. 
The model-independent analysis of these functions impacts 
structure-dependent corrections in hydrogen spectroscopy.%
\footnote{
These functions enter directly in the analysis of spin-dependent transitions. 
Additionally, ansatzes used to model $W_1(0,Q^2)$ may be compared to 
analogous studies of $H_1(0,Q^2)$, which by virtue of an unsubtracted dispersion relation, 
can be reconstructed from experimental data on elastic and inelastic scattering~\cite{Bernard:2007zu}. 
}
Let us compute $H_1(0,Q^2)$ by using NRQED to compute $W_A^{i0}$. This requires a subclass of $1/M^5$ interactions involving the magnetic 
field and four spatial derivatives, 
\begin{align}
\Delta {\cal L} 
&= {g\over M^5} \psi^\dagger \bigg\{ 
c_{Y1} \{ \bm{\sigma}\cdot\bm{B} , \bm{\partial}^4 \} 
+ c_{Y2} \{ \bm{\partial}^2 , \partial^i \bm{\sigma}\cdot\bm{B} \partial^i \} 
+ c_{Y3} \bm{\partial}^2 \bm{\sigma}\cdot\bm{B} \bm{\partial}^2 
+ c_{Y4} \partial^i \partial^j \bm{\sigma}\cdot\bm{B} \partial^i \partial^j 
\nl
&\quad 
+ c_{Y5} ( \bm{B}\cdot\bm{\partial} \bm{\sigma}\cdot\bm{\partial} \bm{\partial}^2  
+ \bm{\partial}^2 \bm{\sigma}\cdot\bm{\partial} \bm{\partial}\cdot \bm{B} )
+ c_{Y6} (\bm{\sigma}\cdot\bm{\partial} \bm{B}\cdot\bm{\partial} \bm{\partial}^2 
+ \bm{\partial}^2 \bm{\partial}\cdot\bm{B} \bm{\sigma}\cdot\bm{\partial} )
\nl
&\quad
+ c_{Y7} ( \partial^i B^j \bm{\sigma}\cdot\bm{\partial} \partial^i \partial^j 
+ \partial^i \partial^j \bm{\sigma}\cdot\bm{\partial} B^j \partial^i )
+ \order(g)  
\bigg\} \psi \,,
\end{align}
where we have neglected effects in setting $\bm{D}\to \bm{\partial}$ that 
are not relevant to the present application. 
At low $Q^2$ we find
\begin{align} \label{eq:H1}
2M H_1(0,Q^2) &= 
{2\over Q^2} Z c_F
+ {1\over 4M^2} \bigg[ Z ( 2c_F + c_S - 2c_{W1} ) - c_F c_D \bigg] 
\nl
& \quad + {Q^2\over 16M^4} \bigg[ c_D(c_{W1} - c_F) - 2Z (c_F + c_{W1} - 32 c_{Y1}) \ +
\nl
& \quad + 32c_F (c_{X1} + c_{X2} + c_{X3}) + 16( - c_{X7} + c_{X8} + c_{X12})
\bigg] + \mathcal{O}(Q^4) .
\end{align} 
Using relations (\ref{eq:c1photon}), (\ref{eq:c1photonXi}), (\ref{eq:virtual}), and
\begin{align} \label{eq:y1photon}
c_{Y1} &= {27\over 256}\bar{F}_1 + {23\over 256} \bar{F}_2 
+ {5\over 16} \bar{F}_1^\prime + {5\over 16} \bar{F}_2^\prime + \frac14 \bar{F}_1^{\prime\prime} 
+ \frac14 \bar{F}_2^{\prime\prime} \,,
\end{align}
the result (\ref{eq:H1}) may be expressed in terms of scattering observables.

\section{Summary \label{sec:summary}}

We have derived a complete basis of operators and coefficient constraints 
through order $1/M^4$ for the parity and time-reversal invariant effective Lagrangian
for a heavy fermion interacting with an abelian gauge field, i.e., NRQED. 

The computations of this paper provide an illustration of relativity 
constraints in high orders of heavy particle 
effective theories~\cite{Heinonen:2012km}.   
In particular, the transformation law for fields under Lorentz boosts receives 
nontrivial corrections compared to a naive reparametrization ansatz~\cite{Luke:1992cs}.  
These effects occur first in the $1/M^4$ Lagrangian, and are validated by 
the explicit matching and variational computations presented in this paper. 
Relations (\ref{eq:constrain4}) and (\ref{eq:fourrelations}), and 
their extensions to QCD (e.g. HQET or NRQCD) represent 
new non-renormalization theorems valid to all orders in perturbation theory.

NRQED can be used to efficiently analyze processes involving 
long wavelength electromagnetic probes of the nucleon.  
The analysis was motivated in part by the necessity to incorporate 
high-order corrections of proton structure in hydrogenic bound states. 
A sample application to structure-dependent
two-photon exchange corrections in the small lepton mass limit
was presented in Section \ref{sec:bound}.  

For applications to low-energy electron-proton scattering, $m_e \ll E \ll m_p$, 
we constructed  in Section \ref{sec:rel} 
the effective theory for a relativistic electron and 
nonrelativistic proton.
The Lagrangian and coefficient relations were derived through $\order(1/M^3)$. 
This case may be applied to a model-independent analysis of 
radiative corrections in the extraction of proton structure from electron 
scattering, as discussed in Section \ref{sec:scattering}.  

The effects of nucleon structure are implemented by
non-pointlike values for the Wilson coefficients 
appearing in (\ref{eq:abelian}).
Having determined the structure of the NRQED Lagrangian, 
it becomes a trivial task to define and compute static properties of nucleons, and 
to unambiguously relate these properties to scattering measurements, 
as illustrated in Section \ref{sec:static}. 

For simplicity we focused in this paper on the effective 
Lagrangian for a parity conserving theory of a heavy fermion coupled to an 
abelian gauge field. 
Extensions to nonabelian gauge fields,  the inclusion of parity violation, 
and the consideration of different heavy particle spins each have important applications.

\vskip 0.2in
\noindent
{\bf Acknowledgements}
\vskip 0.1in
\noindent
Work supported by NSF Grant 0855039. G.~L.~was supported by NSERC.

\end{document}